\magnification \magstep1
\raggedbottom
\openup 2\jot
\voffset6truemm
\def\cstok#1{\leavevmode\thinspace\hbox{\vrule\vtop{\vbox{\hrule\kern1pt
\hbox{\vphantom{\tt/}\thinspace{\tt#1}\thinspace}}
\kern1pt\hrule}\vrule}\thinspace}
\headline={\ifnum\pageno=1\hfill\else
{GREEN FUNCTIONS FOR CLASSICAL EUCLIDEAN MAXWELL THEORY}
\hfill \fi}
\centerline {\bf GREEN FUNCTIONS FOR CLASSICAL}
\centerline {\bf EUCLIDEAN MAXWELL THEORY}
\vskip 1cm
\leftline {GIAMPIERO ESPOSITO}
\vskip 1cm
\noindent
{\it Istituto Nazionale di Fisica Nucleare, Sezione di Napoli,
Mostra d'Oltremare Padiglione 20, 80125 Napoli, Italy}
\vskip 0.3cm
\noindent
{\it Dipartimento di Scienze Fisiche, Mostra d'Oltremare
Padiglione 19, 80125 Napoli, Italy}
\vskip 1cm
\noindent
{\bf Summary.} -
Recent work on the quantization of Maxwell theory has used
a non-covariant class of gauge-averaging functionals
which include explicitly the effects of the extrinsic-curvature
tensor of the boundary, or covariant gauges which, unlike
the Lorentz case, are invariant under conformal rescalings of
the background four-metric. This paper studies in detail the
admissibility of such gauges at the classical level. It is
proved that Euclidean Green functions of a second- or
fourth-order operator exist which ensure the fulfillment of
such gauges at the classical level, i.e. on a portion of
flat Euclidean four-space bounded by three-dimensional surfaces.
The admissibility of the axial and Coulomb gauges is
also proved.
\vskip 3cm
\leftline {PACS numbers: 03.70.+k, 98.80.Hw}
\vskip 100cm
\leftline {\bf 1. - Introduction.}
\vskip 0.3cm
\noindent
The recent progress on Euclidean quantum gravity on manifolds
with boundary [1] has led to a detailed investigation of mixed
boundary conditions for gauge fields and gravitation. These 
have been applied to the one-loop semiclassical evaluation of
quantum amplitudes which are motivated by the analysis of the 
wave function of the universe. In particular, we are here
interested in the quantization programme for Euclidean Maxwell
theory. If one uses Faddeev-Popov formalism, the above amplitudes
involve Gaussian averages over gauge functionals $\Phi(A)$ which
ensure that both the operator on perturbations of $A_{b}$
($A_{b}$ being the electromagnetic potential) and the ghost
operator admit well-defined Feynman Green functions for the
given boundary conditions. As is well known, the gauge-averaging
term ${1\over 2\alpha}[\Phi(A)]^{2}$ ($\alpha$ being a
dimensionless parameter) modifies the second-order operator 
on $A_{b}$ that one would obtain from the Maxwell Lagrangian
${1\over 4}F_{ab}F^{ab}$. For example, the Lorentz choice
$\Phi_{L}(A)=\nabla^{b}A_{b}$ leads to the following 
second-order operator on $A_{b}$ perturbations (the background
value of $A_{b}$ is set to zero):
$$
Q^{bc}=-g^{bc}\cstok{\ }+R^{bc}
+\left(1-{1\over \alpha}\right)\nabla^{b}\nabla^{c} \; .
\eqno (1.1)
$$
With a standard notation, $g$ is the background four-metric,
$\nabla^{b}$ denotes covariant differentiation with respect
to the Levi-Civita connection of the background,
$\cstok{\ }$ is defined as $g^{ab}\nabla_{a}\nabla_{b}$
($-\cstok{\ }$ is the Laplacian having positive spectrum
on compact manifolds), and $R^{bc}$ is the Ricci tensor of
the background. The Feynman choice for $\alpha$: 
$\alpha=1$, reduces $Q^{bc}$ to the standard Hodge-de Rham
operator on one-forms. 

In the underlying {\it classical} theory, one starts from the
purely Maxwell Lagrangian (but bearing in mind that the 
metric is positive-definite, unlike the Lorentzian case),
which is invariant under (infinitesimal) gauge transformations
of the kind
$$
{ }^{f}A_{b}=A_{b}+\nabla_{b}f \; ,
\eqno (1.2)
$$
with the function $f$ being freely specifiable. However, 
when a gauge condition is imposed:
$$
\Phi(A)=0 \; ,
\eqno (1.3)
$$
this leads to non-trivial changes in the original scheme.
They are as follows.
\vskip 0.3cm
\noindent
(i) The preservation of eq. (1.3), viewed as a constraint, has
the effect of turning the original first-class constraints
into the second-class [2].
\vskip 0.3cm
\noindent
(ii) The function $f$ has to obey a differential equation,
instead of being freely specifiable.

For example, suppose one starts from a potential $A_{b}$ 
which does not obey the Lorentz gauge:
$$
\nabla^{b}A_{b} \not = 0 \; .
\eqno (1.4)
$$
Nevertheless, it is possible to ensure that the 
gauge-transformed potential ${ }^{f}A_{b}$ does actually
obey the Lorentz gauge $\nabla^{b}({ }^{f}A_{b})=0$, provided
that $f$ satisfies the second-order equation
$$
\cstok{\ }f=-\nabla^{b}A_{b} \; .
\eqno (1.5)
$$
Denoting by ${\cal G}$ the inverse of the $\cstok{\ }$
operator, i.e. its Green's function, the solution of eq. (1.5)
is formally expressed as
$$
f=-{\cal G} \; \nabla^{b}A_{b} \; .
\eqno (1.6)
$$
For this scheme to hold one has to prove that, with the
given boundary conditions, which are mixed for gauge fields
[1], the inverse of the $\cstok{\ }$ operator actually exists
and can be constructed explicitly. 

For this purpose, sect. {\bf 2} constructs the Green's function
for the non-covariant gauges proposed in refs.[3--5], and sect.
{\bf 3} performs a similar analysis for the conformally invariant
gauge of Eastwood and Singer [6]. 
Sections {\bf 4} and {\bf 5} deal
with the axial and Coulomb gauges, respectively. 
Concluding remarks are presented in sect. {\bf 6}.
\vskip 0.3cm
\leftline {\bf 2. - Green functions with non-covariant gauges.}
\vskip 0.3cm
\noindent
This section provides the classical counterpart of the
analysis in refs.[3--5]. Hence we consider a portion of flat
Euclidean four-space bounded by the three-surfaces $\Sigma_{1}$
and $\Sigma_{2}$, say. The region in between $\Sigma_{1}$
and $\Sigma_{2}$ is foliated by three-dimensional hypersurfaces
with extrinsic-curvature tensor $K$. Denoting by $\beta$ a
dimensionless parameter, the gauges we impose read 
$$
\nabla^{c}A_{c}-\beta A_{0}{\rm Tr}K=0 \; .
\eqno (2.1)
$$
These gauges reduce to the Lorentz choice if $\beta$ is set to
zero, and their consideration is suggested by the need to
characterize the most general class of gauge conditions for
Euclidean Maxwell theory in the presence of boundaries. In this
respect, they are still a particular case of a more general
family of gauges, as shown in ref.[5]. In the quantum theory, they
have the advantage of being the only relativistic gauges for
which the one-loop semiclassical theory can be explicitly
evaluated [5], despite the non-covariant nature resulting from
the $\beta A_{0} \; {\rm Tr}K$ term.  

In the classical theory, the first problem is to make sure that
equation (2.1) can be actually satisfied. This problem was not
even addressed in [3--5], and is the object of our first
investigation. Following the example given in the introduction,
let us assume that the original potential $A_{b}$ (a connection
one-form in geometric language) does not satisfy (2.1). After
the gauge transformation (1.2), the potential 
${ }^{f}A_{b}$ satisfies (2.1) if and only if $f$ obeys the
following equation:
$$
\left(\cstok{\ }- \beta {\rm Tr}K {\partial \over \partial \tau}
\right)f=-\Bigr(\nabla^{c}A_{c}
- \beta A_{0}{\rm Tr}K \Bigr) \; .
\eqno (2.2)
$$
In eq. (2.2) we denote by $\tau$ a radial coordinate, and hence we
are assuming that concentric three-sphere boundaries are studied,
as in [4,5]. In the system of local coordinates appropriate for
the case when flat Euclidean four-space is bounded by concentric
three-spheres, eq. (2.2) takes the form
$$
\left[{\partial^{2}\over \partial \tau^{2}}
+{3(1-\beta)\over \tau}{\partial \over \partial \tau}
+{1\over \tau^{2}} { }^{(3)}\nabla_{i}
{ }^{(3)}\nabla^{i} \right]f=-\Bigr(\nabla^{c}A_{c}
-\beta A_{0}{\rm Tr}K \Bigr) \; ,
\eqno (2.3)
$$
where ${ }^{(3)}\nabla_{i}$ denotes three-dimensional covariant
differentiation tangentially with respect to the Levi-Civita
connection of the boundary. Our first problem is now to find
the Green's function $\widetilde {\cal G}$ 
of the operator in square brackets
in eq. (2.3), so that the function $f$ can be expressed as
$$
f=-{\widetilde {\cal G}} \Bigr(\nabla^{c}A_{c}
-\beta A_{0}{\rm Tr}K \Bigr) \; .
\eqno (2.4)
$$
The general solution of eq. (2.3) is given by the general
solution $f_{0}$ of the homogeneous equation plus a particular
solution of the full equation. The solution $f_{0}$ can 
always be found, since the homogeneous equation is a
second-order differential equation with a regular singular
point at $\tau=0$ (indeed, in our 
two-boundary problem, $\tau$ lies
in the closed interval $[a,b]$, with $a >0$). It is
convenient to study a mode-by-mode form of eq. (2.3), i.e. the
infinite number of second-order equations resulting from the
expansion of $f$ on a family of concentric three-spheres: 
$$
f(x,\tau)=\sum_{n=1}^{\infty}f_{n}(\tau)Q^{(n)}(x) \; ,
\eqno (2.5)
$$
where $Q^{(n)}(x)$ are the scalar harmonics on $S^{3}$, in
the local coordinates $x$. This leads to the equations
$$
\left[{d^{2}\over d\tau^{2}}+{3(1-\beta)\over \tau}
{d\over d\tau}-{(n^{2}-1)\over \tau^{2}}\right]f_{n}
=-\Bigr(\nabla^{c}A_{c}
-\beta A_{0}{\rm Tr}K \Bigr)_{n} \; ,
\eqno (2.6)
$$
for all integer values of $n \geq 1$. The problem of 
inverting a second-order operator and finding $f$ is
therefore reduced to solving for $f_{n}$, for all
$n \geq 1$. For this purpose, it is appropriate to take
an integral transform of both sides of eq. (2.6), and then
anti-transform to obtain $f_{n}$, for all $n \geq 1$. The
integral transform should turn the left-hand side of eq. (2.6)
into the product of a polynomial with the transform of
$f_{n}(\tau)$. To overcome the technical difficulties
resulting from the negative powers of $\tau$ in the
operator, we define a new variable 
$$
w \equiv \log(\tau) \; ,
\eqno (2.7)
$$
so that our problem becomes the one of finding the integral
transform of
$$
\eqalignno{ \; &
\left[{d^{2}\over dw^{2}}+(2-3 \beta){d\over dw}
-(n^{2}-1)\right]f_{n}(w) \cr
&=-{\rm e}^{2w}
\Bigr(\nabla^{c}A_{c}-\beta A_{0}{\rm Tr}K \Bigr)_{n} \; .
&(2.8)\cr}
$$
If we extend the definition of $f_{n}(w)$, requiring that
it should vanish for $w < \log(a)$ and $w > \log(b)$, we
can define its Fourier transform 
$$
{\cal F}(f_{n}(w)) \equiv {\widetilde f}_{n}(p)
\equiv \int_{-\infty}^{\infty}f_{n}(w) 
{\rm e}^{-ip w} dw \; .
\eqno (2.9)
$$
Denoting by $\Omega_{n}$ the right-hand side of eq. (2.8),
one thus finds (hereafter $s \equiv ip$)
$$
f_{n}(w)=\int_{-\infty}^{\infty} 
{{\cal F}(\Omega_{n}){\rm e}^{sw}\over 
[s^{2}+(2-3 \beta)s-(n^{2}-1)]} ds \; ,
\eqno (2.10)
$$
up to a multiplicative constant which is unessential
for our purposes. Of course, the contour in eq. (2.10) has
been rotated to integrate over $s$.
In eq. (2.10), the integrand has poles
corresponding to the zeros of the equation 
$$
s^{2}+(2-3 \beta)s-(n^{2}-1)=0 \; .
\eqno (2.11)
$$
This equation has two real solutions (corresponding to
purely imaginary values of $p$) given by
$$
s={(3\beta -2) \pm \sqrt{(3\beta -2)^{2}+4(n^{2}-1)}
\over 2} \; .
\eqno (2.12)
$$
This means that the integral formula for $f_{n}(w)$ should be
regarded as a contour integration, where the complex contour
goes around the real roots of eq. (2.11). The various possible 
prescriptions are the counterpart of the Green functions of
quantum field theory, i.e. retarded, advanced, Feynman,
Wightman, Hadamard etc. [7]. Note, however, that our analysis
is one-dimensional and entirely classical. The solution of
eq. (2.3) reads therefore
$$
f(x,\tau)=\sum_{i=1}^{2}\sum_{n=1}^{\infty}
a_{n}^{(i)} \; \tau^{\rho_{n}^{(i)}} Q^{(n)}(x)
+\sum_{n=1}^{\infty}f_{n}(w(\tau))Q^{(n)}(x) \; , 
\eqno (2.13)
$$
where $a_{n}^{(i)}$ are constant coefficients and 
$\rho_{n}^{(i)}$ denotes, for $i=1,2$, the roots (2.12).

Interestingly, the value $\beta={2\over 3}$, which was found to
play an important role in the one-loop semiclassical analysis 
of ref.[5], emerges naturally in our case as the particular value
of $\beta$ for which the denominator has real roots
$\pm \sqrt{n^{2}-1}$. When $n=1$, the integrand in eq. (2.10) has
a double pole at $s=0$. This corresponds to the lack of a 
regular decoupled mode for the normal component of the
electromagnetic potential in the one-boundary problem,
in the quantum theory [5].
\vskip 0.3cm
\leftline {\bf 3. - Green's function in the Eastwood-Singer gauge.}
\vskip 0.3cm
\noindent
The gauge conditions studied in the classical theory of the
electromagnetic field are not, in general, invariant under
conformal rescalings of the metric. For example, not even the
Lorentz gauge is preserved under the conformal rescaling 
${\widetilde g}_{ab}=\Omega^{2} \; g_{ab}$ [6]. 
This is not entirely desirable, since we know that the
equations of vacuum Maxwell theory are conformally
invariant. However, as shown in
ref.[6], an operator on one-forms can be defined in such a
way that the resulting gauge condition is conformally invariant.
This reads
$$
\nabla_{b}\biggr[\Bigr(\nabla^{b}\nabla^{c}-2R^{bc}
+{2\over 3}R g^{bc} \Bigr)A_{c}\biggr]=0 \; ,
\eqno (3.1)
$$
where $R$ is the trace of the Ricci tensor. Suppose now that
the original potential $A_{c}$ does not satisfy eq. (3.1). The
gauge-transformed potential obtained according to (1.2) will
instead satisfy eq. (3.1) provided that $f$ obeys the fourth-order
equation
$$
\eqalignno{
\; & \left[\cstok{\ }^{2}+\nabla_{b}\Bigr(-2R^{bc}
+{2\over 3}R g^{bc} \Bigr)\nabla_{c} \right]f \cr
& = -\nabla_{b}\left[\biggr(\nabla^{b}\nabla^{c}
-2R^{bc}+{2\over 3}R g^{bc}\biggr)A_{c}\right] \; ,
&(3.2)\cr}
$$
where $\cstok{\ }^{2}$ denotes the $\cstok{\ }$ operator
composed with itself, i.e. 
$g^{ab}g^{cd}\nabla_{a}\nabla_{b}\nabla_{c}\nabla_{d}$.
Our problem is now to find the Green's function of the
fourth-order operator on the left-hand side of eq. (3.2). Since
our paper is restricted to the analysis of flat Euclidean
backgrounds bounded by concentric three-spheres, we need to
invert the operator [8]
$$ 
\eqalignno{
\cstok{\ }^{2}& \equiv {\partial^{4}\over \partial \tau^{4}}
+{6\over \tau}{\partial^{3}\over \partial \tau^{3}}
+{3\over \tau^{2}} {\partial^{2}\over \partial \tau^{2}}
-{3\over \tau^{3}} {\partial \over \partial \tau} \cr
&+{2\over \tau^{2}}
\left({\partial^{2}\over \partial
\tau^{2}}+{1\over \tau}{\partial \over \partial \tau}
\right)
{ }^{(3)}\nabla_{i}{ }^{(3)}\nabla^{i}
+{1\over \tau^{4}}
{\left({ }^{(3)}\nabla_{i}{ }^{(3)}\nabla^{i}\right)}^{2} \; ,
&(3.3)\cr}
$$
which ensures the fulfillment of the gauge condition
$\cstok{\ }\nabla^{c}A_{c}=0$. As in sect. {\bf 2}, we begin
by taking the Fourier transform of eq. (3.2) when the curvature
of the background vanishes, and a mode-by-mode analysis is
performed. We therefore consider the fourth-order operator [8]
$$
\cstok{\ }_{n}^{2} \equiv  
{d^{4}\over d\tau^{4}}+{6\over \tau}{d^{3}\over d\tau^{3}}
-{(2n^{2}-5)\over \tau^{2}}{d^{2}\over d\tau^{2}} 
-{(2n^{2}+1)\over \tau^{3}}{d\over d\tau}
+{(n^{2}-1)^{2}\over \tau^{4}} \; .
\eqno (3.4) 
$$
The introduction of the variable (2.7) leads to a remarkable
cancellation of some derivatives. Hence one finds the
equation
$$
\eqalignno{ \; & 
\left[{d^{4}\over dw^{4}}-2(n^{2}+1){d^{2}\over dw^{2}}
+(n^{2}-1)^{2}\right]f_{n}(w) \cr
&=-{\rm e}^{4w}\Bigr(\cstok{\ }
\nabla^{c}A_{c}\Bigr)_{n} \; .
&(3.5)\cr}
$$
We now take the Fourier transform of eq. (3.5), and then
anti-transform setting $s \equiv ip$. This leads 
to the contour formula (cf. eq. (2.10))
$$
f_{n}(w) = \int_{-\infty}^{\infty}
{{\cal F}({\widetilde \Omega}_{n}){\rm e}^{sw}
\over [s^{4}-2(n^{2}+1)s^{2}
+(n^{2}-1)^{2}]} ds \; ,
\eqno (3.6)
$$
where ${\widetilde \Omega}_{n}$ is the right-hand side of
eq. (3.5). The structure of the poles of the integrand differs
substantially from the one found in sect. {\bf 2}.
The poles are now the roots of the fourth-order algebraic
equation
$$
s^{4}-2(n^{2}+1)s^{2}+(n^{2}-1)^{2}=0 \; ,
\eqno (3.7)
$$
which has the four real roots $\pm (n \pm 1)$ [8].

In our flat background with boundary, the general solution of
eq. (3.2) is then given by 
$$
f(x,\tau)=u(x,\tau)+\sum_{n=1}^{\infty}f_{n}(w(\tau))
Q^{(n)}(x) \; ,
\eqno (3.8)
$$
where $u(x,\tau)$ solves the homogeneous equation, and can
be written as
$$
u(x,\tau)=\sum_{i=1}^{4} \sum_{n=1}^{\infty}
b_{n}^{(i)} \; \tau^{\sigma_{n}^{(i)}} \;
Q^{(n)}(x) \; ,
\eqno (3.9)
$$
where $\sigma_{n}^{(i)}$ denotes, for $i=1,2,3,4$, a solution
of eq. (3.7), and $b_{n}^{(i)}$ are constant coefficients.
\vskip 0.3cm
\leftline {\bf 4. - Axial gauge.}
\vskip 0.3cm
\noindent
Covariant gauges are not the only possible choice for the
quantization of Maxwell theory. By contrast, non-covariant
gauges play an important role as well, and a comprehensive
review may be found in ref.[9]. In our paper we focus on the
axial and Coulomb gauges at the classical level. Let $n^{a}$
be the normal to the hypersurface $\Sigma$ belonging to the
family of hypersurfaces foliating the portion of flat
Euclidean four-space bounded by $\Sigma_{1}$ and $\Sigma_{2}$. 
If $A_{b}$ does not obey the axial gauge:
$$
n^{b}A_{b} \not = 0 \; ,
\eqno (4.1)
$$
one can nevertheless perform the transformation (1.2)
and then look for a function $f$ such that
$$
n^{b}(A_{b}+\nabla_{b}f)=0 \; .
\eqno (4.2)
$$
We can safely assume that $n^{a}$ takes the form 
$(1,0,0,0)$, while $\Sigma_{1}$ and $\Sigma_{2}$
consist (once again) of concentric three-spheres. In such
a case, eq. (4.2) reduces to
$$
{\partial f \over \partial \tau}=-A_{0}(x,\tau) \; ,
\eqno (4.3)
$$
which can be solved for $f$ in the form
$$
f(x,\tau)=f(x,a)-\int_{a}^{\tau}A_{0}(x,y)dy \; ,
\eqno (4.4)
$$
for all $\tau \in [a,b]$, where $a$ and $b$ are the radii
of the three-spheres $\Sigma_{1}$ and $\Sigma_{2}$, respectively.
\vskip 10cm
\leftline {\bf 5. - Coulomb gauge.}
\vskip 0.3cm
\noindent
The Coulomb gauge is a non-covariant gauge which involves 
the three-dimensional divergence of the 
three-dimensional vector potential [1]. By
this we mean that tangential covariant derivatives of 
$A_{i}$ are taken with respect to the Levi-Civita connection
of the induced metric on the boundary. These three-dimensional
covariant derivatives are denoted by ${ }^{(3)}\nabla_{i}$ 
in our paper, and more often by $\mid_{i}$ in the literature
on general relativity. If $A_{i}$ does not satisfy the
Coulomb gauge, we consider the tangential components of the
gauge transformation (1.2), and we require that
$$
{ }^{(3)}\nabla^{i} \Bigr(A_{i}+{ }^{(3)}\nabla_{i}f
\Bigr)=0 \; .
\eqno (5.1)
$$
The tangential components $A_{i}$ admit a standard expansion 
on a family of concentric three-spheres according to [1]
$$
A_{i}(x,\tau)=\sum_{n=2}^{\infty}
\Bigr[\omega_{n}(\tau)S_{k}^{(n)}(x)
+g_{n}(\tau)P_{k}^{(n)}(x)\Bigr] \; ,
\eqno (5.2)
$$
where $S_{k}^{(n)}(x)$ and $P_{k}^{(n)}(x)$ are
transverse and longitudinal harmonics on $S^{3}$,
respectively [1]. Bearing in mind that [1]
$$
S_{k}^{(n) \; \mid k}(x)=0 \; ,
\eqno (5.3)
$$
$$
P_{k}^{(n) \; \mid k}(x)=-Q^{(n)}(x) \; ,
\eqno (5.4)
$$
eq. (5.1) takes the form
$$
\sum_{n=2}^{\infty}\Bigr[(n^{2}-1)f_{n}(\tau)
+g_{n}(\tau)\Bigr]Q^{(n)}(x)=0 \; ,
\eqno (5.5)
$$
which implies that
$$
f_{n}(\tau)=-{g_{n}(\tau)\over (n^{2}-1)}
\; \; {\rm for} \; {\rm all} \; n \geq 2 \; ,
\eqno (5.6)
$$
while $f_{1}(\tau)$ remains freely specifiable. The classical
{\it gauge} mode $f_{1}$ is not fixed by the condition (5.1),
since its coefficient vanishes (it equals $n^{2}-1$ evaluated
at $n=1$). The result (5.6) can be now 
inserted into the expansion (2.5)
to express the gauge function as an infinite sum of modes
proportional to longitudinal modes. Hence the $f_{n}$ modes
obey the same boundary conditions imposed on longitudinal
modes of $A_{i}$. The residual gauge freedom encoded by
$f_{1}(\tau)$, however, deserves further investigation (cf. 
the quantum analysis in section 7.8 of ref.[1]).
\vskip 0.3cm
\leftline {\bf 6. - Concluding remarks.}
\vskip 0.3cm
\noindent
Our paper has studied the admissibility of non-covariant
or conformally invariant gauge conditions for classical
Maxwell theory in the presence of boundaries. Since the
background is flat Euclidean (instead of flat Minkowskian),
our analysis provides the classical counterpart of the
Euclidean version of the quantum theory. We might have 
addressed the simpler problem of imposing that both the
original potential and the gauge-transformed one (see
(1.2)) satisfy the gauge condition.
We have instead studied the more difficult
case, where the gauge condition is eventually fulfilled after
performing the transformation (1.2). The function $f$ is
no longer freely specifiable, but obeys second- or
fourth-order equations, whose Green's functions have been
explicitly evaluated in sects. {\bf 2} 
and {\bf 3}. Moreover, sects. {\bf 4}
and {\bf 5} show that, when the axial or Coulomb gauge are
imposed, it is simpler to solve for the gauge function $f$.
Equation (4.4) provides an integral representation, while
(2.5) and (5.6) yield an infinite sum for $f$, with modes
proportional to the longitudinal modes for $A_{i}$.

Our analysis puts on solid ground the consideration of the
gauge conditions (2.1) and (3.1) for classical Euclidean Maxwell 
theory on manifolds with boundary. The consideration of boundary
effects is of crucial importance for a correct formulation
of the classical and quantum theories: potential theory,
Casimir effect and one-loop quantum cosmology are some of
the many examples one may provide [1]. It now remains to be
seen how to include the effects of curvature (of the
background) in the evaluation of our Green's functions. 
Moreover, a deep problems consists in studying the possible
occurrence of Gribov phenomena if eqs. (2.1) and (3.1) are applied
to the quantization of non-Abelian gauge theories [10].

On the quantum side, further developments are likely to
occur in the analysis of the one-loop effective action
in non-covariant [11] or covariant [12] gauges, and in
the application of such techniques to the evaluation of
Casimir energies [13]. If such a programme could be 
accomplished, it would provide further evidence in favour
of quantum cosmology having a deep influence on current
developments in the theory of quantized gauge fields [1].
\vskip 0.3cm
\centerline {* * *}
\vskip 0.3cm
The author is indebted to Alexander Kamenshchik, Giuseppe
Pollifrone and Klaus Kirsten for scientific collaboration
on Euclidean Maxwell theory. Many thanks are also due to
Ivan Avramidi and Dima Vassilevich for enlightening
conversations.
\vskip 0.3cm
\leftline {REFERENCES}
\vskip 0.3cm
\item {[1]}
ESPOSITO G., KAMENSHCHIK A. Yu. and POLLIFRONE G.,
{\it Euclidean Quantum Gravity on Manifolds with Boundary,
Fundamental Theories of Physics}
(Kluwer, Dordrecht) 1997.
\item {[2]}
HANSON A., REGGE T. and TEITELBOIM C., {\it Constrained
Hamiltonian Systems} (Accademia dei Lincei, Rome) 1976.
\item {[3]}
ESPOSITO G., {\it Class. Quantum Grav.}, {\bf 11} (1994) 905.
\item {[4]}
ESPOSITO G., KAMENSHCHIK A. Yu., MISHAKOV I. V. and
POLLIFRONE G., {\it Class. Quantum Grav.}, {\bf 11}
(1994) 2939.
\item {[5]}
ESPOSITO G., KAMENSHCHIK A. Yu., MISHAKOV I. V. and
POLLIFRONE G., {\it Phys. Rev. D}, {\bf 52} (1995) 2183.
\item {[6]}
EASTWOOD M. and SINGER M., {\it Phys. Lett. A}, 
{\bf 107} (1985) 73.
\item {[7]}
FULLING S. A., {\it Aspects of Quantum Field Theory in
Curved Spacetime} (Cambridge University Press, Cambridge) 1989.
\item {[8]}
ESPOSITO G., {\it Quantized Maxwell Theory in a Conformally
Invariant Gauge} (DSF preprint 96/42, HEP-TH 9610017).
\item {[9]}
LEIBBRANDT G., {\it Rev. Mod. Phys.}, {\bf 59}
(1987) 1067.
\item {[10]}
GRIBOV V. N., {\it Nucl. Phys. B}, {\bf 139} (1978) 1.
\item {[11]}
ESPOSITO G., KAMENSHCHIK A. Yu. and POLLIFRONE G.,
{\it Class. Quantum Grav.}, {\bf 13} (1996) 943.
\item {[12]}
ESPOSITO G., KAMENSHCHIK A. Yu. and KIRSTEN K.,
{\it Phys. Rev. D}, {\bf 54} (1996) 7328.
\item {[13]}
LESEDUARTE S. and ROMEO A., {\it Ann. Phys. (N.Y.)},
{\bf 250} (1996) 448.
 
\bye